\documentclass[aps, prl, reprint, nofootinbib, superscriptaddress, showpacs]{revtex4-1}
\usepackage{graphicx}
\usepackage{dcolumn}
\usepackage{bm}
\usepackage{color}
\usepackage{amsmath}
\usepackage{amssymb}
\usepackage{graphicx}% Include figure files
\usepackage{url}

\begin{document}
\preprint{APS/123-QED}

\title{Photoelectron holographic interferometry to probe the longitudinal momentum offset at the tunnel exit}

\author{Min Li}
\email{mli@hust.edu.cn}
\author{Hui Xie}
\author{Wei Cao}
\author{Siqiang Luo}
\author{Jia Tan}
\author{Yudi Feng}
\author{Baojie Du}
\author{Weiyu Zhang}
\author{Yang Li}
\author{Qingbin Zhang}
\author{Pengfei Lan}
\author{Yueming Zhou}
\email{zhouymhust@hust.edu.cn}

\affiliation{Wuhan National Laboratory for Optoelectronics and School of Physics, Huazhong University of Science and Technology, Wuhan 430074, China}

\author{Peixiang Lu}
\email{lupeixiang@hust.edu.cn}
\affiliation{Wuhan National Laboratory for Optoelectronics and School of Physics, Huazhong University of Science and Technology, Wuhan 430074, China} 
\affiliation{Hubei Key Laboratory of Optical Information and Pattern Recognition, Wuhan Institute of Technology, Wuhan 430205, China}

\date{\today}

\begin{abstract}
Laser-induced electron tunneling underlies numerous emerging spectroscopic techniques to probe attosecond electron dynamics in atoms and molecules. The improvement of those techniques requires an accurate knowledge of the exit momentum for the tunneling wave packet. Here we demonstrate a photoelectron interferometric scheme to probe the electron momentum longitudinal to the tunnel direction at the tunnel exit by measuring the photoelectron holographic pattern in an orthogonally polarized two-color laser pulse. In this scheme, we use a perturbative 400-nm laser field to modulate the photoelectron holographic fringes generated by a strong 800-nm pulse. The fringe shift offers a direct experimental access to the intermediate canonical momentum of the rescattering electron, allowing us to reconstruct the momentum offset at the tunnel exit with high accuracy. Our result unambiguously proves the existence of nonzero initial longitudinal momentum at the tunnel exit and provides fundamental insights into the non-quasi-static nature of the strong-field tunneling.
\end{abstract}

\maketitle

The tunneling of an electron through the suppressed Coulomb potential barrier is one of the most fundamental processes in strong-field light-matter interactions \cite{Keldysh}. Since strong-field tunneling initiates plenty of intriguing phenomena, such as high-harmonic generation and nonsequential double ionization, the electron initial momentum at the tunnel exit is essential for a deepened understanding of those subsequent processes. Generally, the initial momentum at the tunnel exit is small. Starting with this initial momentum, the tunneling electron is accelerated by the laser field and achieves a large laser-induced drift momentum on its subsequent motion. The smaller initial momentum is easily buried below the larger laser-induced drift momentum, thus it is difficult to directly resolve the initial momentum at the tunnel exit from the experiment.   

Much of previous attention has concentrated on the initial momentum transverse to the tunnel direction \cite{Perelomov,VDMur,Barth,Eckart,Ohmi}. The momentum component longitudinal to the instantaneous tunnel direction (initial longitudinal momentum) is more difficult to identify because the electron momentum along this direction is continuously changed by the laser electric field \cite{MYIvanov}. Thus the initial longitudinal momentum at the tunnel exit is intertwined with the instant of tunneling \cite{Mli_pra2016}. Usually, the initial longitudinal momentum is assumed to be zero in the adiabatic picture \cite{Ammosov,Delone}, where the tunneling is treated as if the electron penetrates a static barrier. This assumption is widely used by many versions of the classical or semiclassical models \cite{Corkum1993, Paulus,Hu,Mli_2014}. However, some recent experiments \cite{Camus,Pfeiffer,Hofmann,XSun_pra2014,Pedatzur} and simulations \cite{Tian,Xu,Teeny} have shown that this assumption might be inaccurate. In particular, the width of the initial longitudinal momentum spread at the tunnel exit was estimated by comparing the measured photoelectron momentum distributions with semiclassical simulations \cite{Pfeiffer,Hofmann,XSun_pra2014}. With assuming that the ionization and rescattering times are known from theories, a nonzero initial longitudinal momentum offset was revealed for different harmonic orders \cite{Pedatzur}. Using the backpropagation method, it was shown that the initial longitudinal momentum has a significant effect in retrieving the tunneling exit information \cite{Ni1}. Thus knowledge about the initial longitudinal momentum is essential for the understanding and controlling of the tunneling ionization. Up to now, the value of the initial longitudinal momentum at the tunnel exit is still under debate (see  \cite{MYIvanov,Mli_pra2016,Tian,Xu,Teeny} for the predictions by different models).

In this Letter, we propose and demonstrate an experimental scheme to precisely detect the initial longitudinal momentum at the tunnel exit using strong-field photoelectron holography (SFPH) \cite{Huismans_sci2011}. The key of this scheme is to experimentally extract the intermediate canonical momentum of the rescattering electron, a quantity that bridges the initial tunneling step and the subsequent scattering process. In strong-field approximation theory \cite{Salieres,Becker_2002}, the rescattering electron is initially released with the intermediate canonical momentum (or wave vector) $\textbf{k}$ and then scatters into the final momentum $\textbf{p}$. We show that the intermediate canonical momentum $\textbf{k}$ of the rescattering electron is encoded in the fringe shift of the SFPH. Due to the rescattering process, the longitudinal and transverse components of the intermediate canonical momentum are coupled with each other, allowing us to reconstruct the initial longitudinal momentum offset from their coupling relation. Our result provides unambiguous evidence for the existence of nonzero initial longitudinal momentum at the tunnel exit, which contradicts the common assumption in the adiabatic picture \cite{Ammosov,Delone}. 
 
The SFPH is a powerful ultrafast photoelectron spectroscopy of electron and nuclear dynamics in atoms and molecules \cite{Bian1, Bian2,Meckel,MMLiu_prl2016,YZhou_prl2016,Walt,MHe_prl2018,JTan_oe2018,Porat}. In SFPH, the interference of the signal (rescattering electron) and reference (direct electron) waves manifests itself as a spider-like structure in the photoelectron momentum distribution. Our experimental scheme is based on the SFPH in an orthogonally polarized two-color (OTC) laser pulse \cite{Ivanov,Kitzler,LZhang,Richter,XGong}, which is given by,
\begin{eqnarray}
\textbf{E}(t)=E_{z,800} \cos(\omega t)\textbf{e}_z +E_{y,400} \cos(2\omega t+\varphi)\textbf{e}_y.
\end{eqnarray}
Here $\omega$ is the frequency of the 800-nm field, and $\varphi$ is the relative phase between the two-color components. Our experiment utilizes a strong 800-nm fundamental wave (FW) ($\sim 6.7\times10^{13} \rm{W/cm^2}$) and a perturbative 400-nm second harmonic (SH) pulse ($\sim 1.4\times10^{12} \rm{W/cm^2}$). In the present study, $\textbf{e}_z$ and $\textbf{e}_y$ are referred to as longitudinal and transverse directions, respectively. Note the longitudinal direction is different from the definition in recent work on photon momentum sharing between electrons and nuclei \cite{Chelkowski}. The three-dimensional photoelectron momentum distributions from strong-field ionization of Ar are measured using a cold target recoil ion momentum spectroscopy (COLTRIMS) \cite{Ullrich}. Further experimental details are given in \cite{supplemetary}. Atomic units (a.u.) are used throughout unless stated otherwise.

Figures 1(a) and 1(b) show the measured photoelectron momentum distributions at two typical relative phases of the OTC fields. The characteristic spider-like structures originating from the SFPH are visible. Those spider-like structures are not symmetric with $p_y=0$ in the OTC fields. The central maxima (guided by the dashed lines at near $p_y=0$) of the holographic pattern shift to negative $p_y$ at the relative phase in Fig.\,1(a) and to positive $p_y$ at the relative phase in Fig.\,1(b). To see the fringe shift more clearly, we show in Fig.\,1(c) the measured $p_y$ momentum distribution with respect to the relative phase $\varphi$ at $p_z=0.4$ a.u. One can see that the central maximum (guided by the black solid line) oscillates with $\varphi$ with an amplitude of $\sim$0.045 a.u. along the transverse direction. We have also numerically solved the time-dependent Schr\"{o}dinger equation (TDSE) of Ar in the OTC fields \cite{supplemetary}. The TDSE result shown in Fig.\,1(d) agrees with the experimental results for the oscillation of the central maximum with $\varphi$. Comparing the TDSE result with the experiment, we can calibrate the relative phase between the two-color components. The TDSE result reveals some minima of the photoelectron yield in the central maximum corresponding to the largest electron initial transverse momentum \cite{JTan_oe2018}, which are blurred in the experiment probably due to the focal volume effect.

\begin{figure}
	\includegraphics[width=8.5cm]{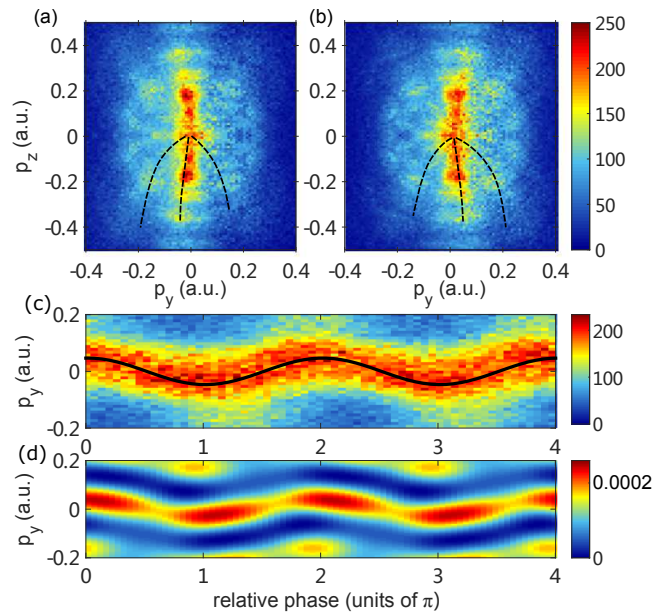}
	\caption{\label{fig1} 
		(a) and (b) display the measured photoelectron momentum distributions of Ar atom in OTC fields at two relative phases, corresponding to the most asymmetric fringes for the SFPH. The black dashed lines indicate the maxima of photoelectron holographic fringes. (c) and (d) show the measured and TDSE simulated photoelectron $p_y$ momentum distributions as a function of the relative phase at $p_z=0.4$ a.u., respectively. The solid black line in (c) shows the shift of the central maxima as a function of $\varphi$. The experimental data are integrated over a momentum range $\left| p_x\right| <0.1$ a.u., where $p_x$ is the electron momentum along the laser propagation direction.
	}	 
\end{figure}

The pattern of the SFPH is determined by the phase difference between the direct and rescattering electron wave packets. In a single-color field, the phase difference is expressed as \cite{YZhou_prl2016,JTan_oe2018} $\Delta S\approx\frac{p_y^2}{2}(t_c-t_0)+\alpha$, where $p_y$ is the final momentum perpendicular to the polarization direction, $t_c$ is the rescattering time, $t_0$ is the ionization time, and $\alpha$ is the phase arising from the interaction between the electron and the parent ion. Adding a perturbative SH field polarized perpendicular to the strong FW, the ionization and rescattering times are unchanged while the phase difference between the signal and reference waves is changed to \cite{supplemetary},
\begin{eqnarray}
\Delta S\approx \frac{(p_y-k_y)^2}{2}(t_c-t_0)+\alpha,
\end{eqnarray} 
where $k_{y}=-\frac{1}{t_c-t_s}\int_{t_s}^{t_c}A_{y}(\tau)d\tau$ is the intermediate canonical momentum of the rescattering electron along the SH direction, $t_s$ is the saddle-point time, and $A(\tau)$ is the laser vector potential. Because of the perturbative nature of the SH field, the phase $\alpha$ is nearly the same for the single-color and OTC fields. Comparing the phase difference in Eq.\,(2) with that in single-color field, we obtain that the shift of the central maximum with respect to $p_y=0$ (the central maximum appears at $p_y=0$ in the single-color field) is the transverse component of the intermediate canonical momentum $k_y$ in the OTC field.

To validate this result, we show in Fig.\,2(a) the measured shifts of the central maxima of the holographic pattern as functions of $p_z$ and $\varphi$. For each $p_z$, the shift of the holographic fringes oscillates with $\varphi$. The amplitude of the oscillation slightly increases with $p_z$, as shown by the color scale. The corresponding TDSE result shown in Fig.\,2(b) is consistent with the experiment. Figure 2(c) shows the calculated $k_y$ by numerically solving the saddle-point equation for the rescattering electron. One can see that the calculated $k_y$ by solving the saddle-point equation is nearly the same as the fringe shifts in both experiment and TDSE, which demonstrates that the fringe shift corresponds to the transverse component $k_y$ of the intermediate canonical momentum. Moreover, for the near-forward scattering electron, the longitudinal component of the intermediate canonical momentum $k_z$ is nearly unchanged at the instant of rescattering and thus it approximately equals the measured final longitudinal momentum $p_z$, i.e., $k_z\approx p_z$. Thus the vector of the intermediate canonical momentum $\textbf{k}$ of the rescattering electron has been extracted.

The correspondence between the fringe shift of the SFPH and the intermediate canonical momentum can also be explained remarkably simply. The central maximum of the holographic pattern corresponds to zero scattering angle at the instant of rescattering. As a result, the intermediate canonical momentum of the rescattering electron is unchanged during the scattering process. Therefore, the final momentum vector of the central maximum is exactly the intermediate canonical momentum of the rescattering electron.

\begin{figure}
	\includegraphics[width=9.0cm]{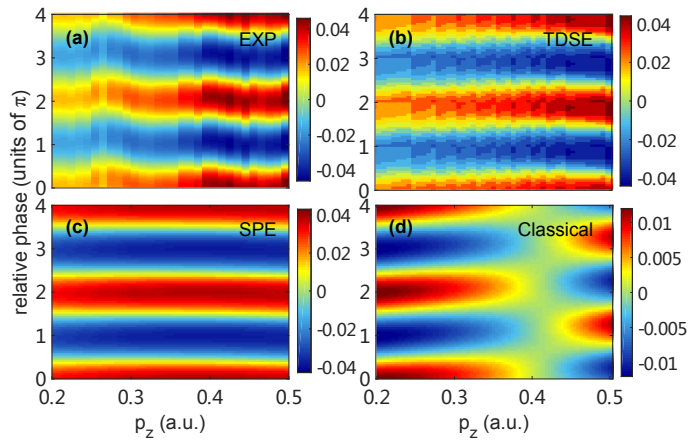}
	\caption{\label{fig2} 
		(a) and (b) show the shifts of the central maxima of the holographic patterns with respect to $p_z$ and $\varphi$ from the experiment and TDSE, respectively. (c) and (d) show the calculated intermediate canonical momentum $k_y$ by solving the saddle-point equation (SPE) and by the classical rescattering model with assuming zero initial longitudinal momentum, respectively.
	}	 
\end{figure}

The intermediate canonical momentum \textbf{k} is directly linked to the initial momentum at the tunnel exit \textbf{v} by $\textbf{k}=\textbf{v}-\textbf{A}(t_0)$. So it is possible to reconstruct the initial longitudinal momentum offset from the extracted intermediate canonical momentum. To do this, one should first determine the ionization time $t_0$ of the tunneling electron with sufficient precision \cite{Pedatzur,Shafir}, which is generally difficult in experiments. This difficulty can be overcome in our scheme. Because the rescattering electron is driven back to the parent ion along both longitudinal and transverse directions (the rescattering condition), the longitudinal component $k_z$ of the intermediate canonical momentum are coupled with its transverse component $k_y$. Both of them are functions of the initial longitudinal momentum $v_z$. Since both $k_z$ and $k_y$ are direct experimental observables, we can obtain the initial longitudinal momentum at the tunnel exit by resolving the coupling problem between $k_z$ and $k_y$ without any assumption on the ionization time. 

To reveal the sensitivity of the transverse component $k_y$ of the intermediate canonical momentum on the initial longitudinal momentum $v_z$, we show in Fig.\,2(d) $k_y$ calculated by the classical rescattering model \cite{Corkum1993,Paulus}, in which zero initial longitudinal momentum is assumed at the tunnel exit \cite{supplemetary}. The classically calculated $k_y$ exhibits two striking differences from the measured fringe shift in Fig.\,2(a). The first one is that the amplitude of the oscillation for the classically calculated $k_y$ (the color scale) is much smaller than that of the measured fringe shift. Figure 3(a) shows the lineouts taken from Figs.\,2(a) and 2(d) at $p_z=0.4$ a.u. The amplitude for the experimental result is $\sim$0.045 a.u., while for the classically calculated $k_y$ (blue solid line) is only $\sim$0.001 a.u. The other obvious difference is the phase jump at $p_z\simeq0.43$ a.u. for the classical result, as shown in Fig.\,2(d), which does not appear in the experiment. The significant difference between the experiment and the classical rescattering model shown in Fig.\,2 implies that the intermediate canonical momentum is very sensitive to the initial longitudinal momentum at the tunnel exit. Thus we can reconstruct the initial longitudinal momentum offset from the extracted intermediate canonical momentum with high accuracy.

%Reconstruction of the momentum when an electron exits the tunneling barrier
In the reconstruction procedure \cite{supplemetary}, we assume an arbitrary $v_z$ for a specific $k_z$ (or, equivalently, $p_z$), and then calculate the transverse component $k_y^{\mathrm{cal}}(v_z)$ using the classical rescattering model. By scanning $v_z$, the initial longitudinal momentum offset can be obtained if the calculated $k_y^{\mathrm{cal}}(v_z)$ agrees with the measured fringe shift. We calculate the standard deviation $\sigma(v_z)=\frac{1}{N}\sqrt{\sum_{n=1}^{N}[k_y^{\mathrm{cal}}(v_z)-k_y^{\mathrm{exp}}]^2}$ as a function of $v_z$, where $k_y^{\mathrm{exp}}$ is the extracted $k_y$ from the measurement and $N$ is the number of the relative phases. The initial longitudinal momentum offset is reconstructed by minimizing the standard deviation $\sigma(v_z)$. Figure 3(b) shows the standard deviation with respect to $v_z$ for $p_z=0.4$ a.u. A minimum of the standard deviation appears at $v_z=0.189$ a.u. With this initial longitudinal momentum offset included in the classical rescattering model, we can achieve a good agreement with the experiment, as shown by the red dashed lines in Fig.\,3(a).

\begin{figure}
	\includegraphics[width=8.5cm]{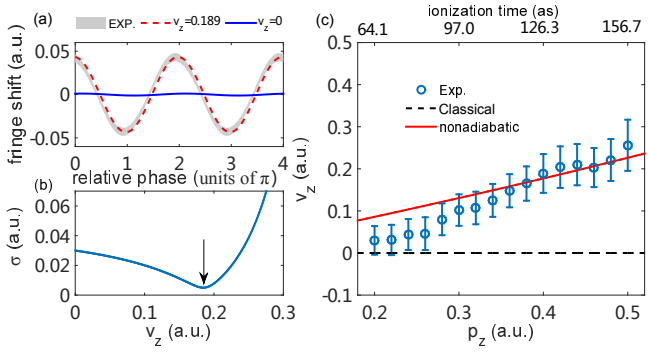}
	\caption{\label{fig3} 
		(a) The measured fringe shift (gray area) with respect to $\varphi$ at $p_z=0.4$ a.u. The blue solid (red dashed) lines show the classical predictions with zero initial longitudinal momentum (with an initial longitudinal momentum of  $v_z=0.189$ a.u.). The uncertainty of the measured fringe shift is included within the gray area. (b) The standard deviation $\sigma(v_z)$ with respect to the initial longitudinal momentum for $p_z=0.4$ a.u.. The arrow indicates the minimum of the standard deviation. (c) The reconstructed initial longitudinal momentum offset $v_z$ with respect to the final momentum $p_z$. The predictions by the subcycle nonadiabatic theory \cite{Mli_pra2016} and the classical rescattering model are shown by the red solid and black dashed lines, respectively.
	}	 
\end{figure}

The reconstructed initial longitudinal momentum offset as a function of $p_z$ is shown in Fig.\,3(c). One can see that the reconstructed initial longitudinal momentum is approximately linear with the final momentum $p_z$. With increasing $p_z$, the initial longitudinal momentum offset $v_z$ increases. This is in contrast to the adiabatic tunneling theory \cite{Ammosov,Delone}, in which zero initial longitudinal momentum is assumed. This observation is consistent with previous experiment using high harmonic spectroscopy \cite{Pedatzur}, and the reconstructed initial longitudinal momentum offset agrees well with the prediction of the subcycle nonadiabatic tunneling theory \cite{Mli_pra2016}, as shown by the red solid line in Fig.\,3(c). 

Since the rescattering electron travels along a two-dimensional trajectory in the OTC field, it needs a nonzero initial transverse momentum $v_y$ at the tunnel exit to compensate the electron motion induced by the laser field along the SH direction. This initial transverse momentum can also be reconstructed. The initial transverse momentum for the rescattering electron is directly linked with $k_y$ by $v_y=k_y+A_y(t_0,\varphi)$, where $t_0$ is given by $t_0=\frac{1}{\omega}\sin^{-1}\frac{\omega(k_z-v_z)}{E_{z,800}}$. Thus, the initial transverse momentum is not only a function of $p_z$, but also a function of $\varphi$. Figure 4(a) shows the reconstructed initial transverse momentum $v_y$ with respect to $p_z$ and $\varphi$. The predictions by solving the saddle-point equation and the classical rescattering model are shown in Figs.\,4(b) and 4(c), respectively. As guided by the white dots, the maximal $v_y$ in the experiment and the saddle-point equation is nearly unchanged with increasing $p_z$, while it decreases for the classical rescattering model. This difference is more clearly seen by comparing the lineouts taken from Figs.\,4(a)-4(c) at $p_z=0.2$ a.u. [Fig.\,4(d)], $p_z=0.3$ a.u. [Fig.\,4(e)], and $p_z=0.4$ a.u. [Fig.\,4(f)]. At small $p_z$ [Fig.\,4(d)], the classical result agrees with the experiment and TDSE. With increasing $p_z$, the classical result deviates largely with the experiment and TDSE, as shown in Figs.\,4(e) and 4(f). Because of the correspondence between the ionization time and the final momentum $p_z$ [Fig.\,3(c)], our result demonstrates that nonadiabatic effects become significant when the electron tunnels at a laser phase away from the laser field crest \cite{Mli_pra2016}. 

\begin{figure}
	\includegraphics[width=8.5cm]{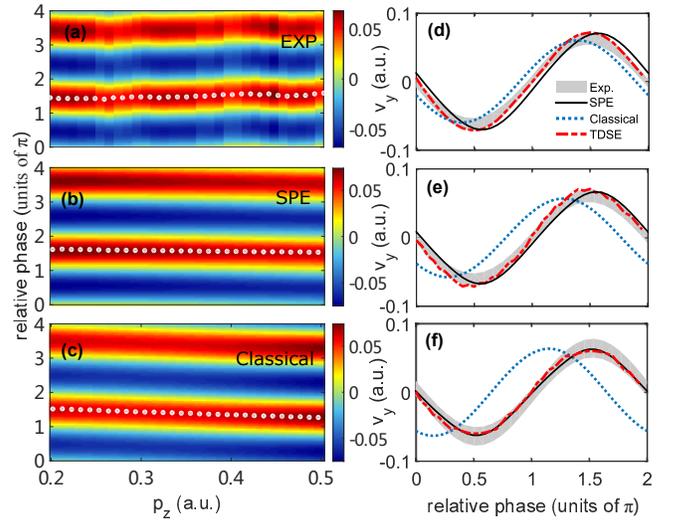}
	\caption{\label{fig4} 
		(a) shows the reconstructed initial transverse momentum offset $v_y$ of the rescattering electron at the tunnel exit with respect to $p_z$ and $\varphi$ from the experiment. (b) and (c) show the predicted $v_y$ by solving the saddle-point equation (SPE) and the classical rescattering model, respectively. The relative phase corresponding to the maximal $v_y$ is marked by the white dots. (d)-(f) show the initial transverse momentum offset $v_y$ with respect to $\varphi$ at $p_z=0.2$ a.u., $p_z=0.3$ a.u., and $p_z=0.4$ a.u., respectively. The uncertainty of the experiment is included within the gray area.
	}	 
\end{figure}
%Discussion 
It has been shown before that the Coulomb potential has a significant influence on the SFPH \cite{Huismans_sci2011,HXie_oe2016}. Since the 400-nm field is weak enough, the phase $\alpha$ arising from the Coulomb potential in Eq.\,(2) is nearly the same for the single-color and OTC fields. As a result, the Coulomb effect has been mechanically excluded by comparing the central maximum of the holographic pattern in the OTC field with that in the single-color field \cite{supplemetary}, which is one of the most distinctive advantages of this reconstruction method. This is also confirmed by the agreement between the measured fringe shift and the $k_y$ calculated by the saddle-point equation without consideration of the Coulomb potential, as shown in Fig. 2. However, when $p_z$ is within [0.2, 0.28] a.u., the amplitude for the oscillation of the fringe shifts in experiment [Fig.\,2(a)] and TDSE [Fig.\,2(b)] is a little smaller than that of the $k_y$ calculated by the saddle-point equation [Fig.\,2(c)]. This comes from the large difference of the Coulomb effect on the SFPH between the single-color and OTC fields when $p_z$ approaches zero \cite{HXie_oe2016}. At this momentum range, the phase $\alpha$ in Eq.\,(2) is different for the single-color and OTC fields. This also leads to the small differences between the reconstructed initial longitudinal momentum offset and the prediction of the nonadiabatic tunneling theory when $p_z$ is within [0.2, 0.28] a.u., as shown in Fig.\,3(c). 

%Extending our method to longer laser wavelength can diminish the Coulomb effect and can increase the momentum range of the SFPH \cite{MHe_prl2018,JTan_oe2018}.

In summary, we have measured the photoelectron holographic patterns of Ar atom in OTC laser fields. We demonstrate that the intermediate canonical momentum of the rescattering electron can be extracted from the fringe shift of the photoelectron holographic structure. Because the intermediate canonical momentum is very sensitive to the initial condition set by the tunneling ionization, we can reconstruct the electron initial longitudinal momentum offset at the tunnel exit with high accuracy. Our experiment demonstrates that the tunneling electron is released with nonzero initial momentum longitudinal to the tunnel direction. This does not concur with the fundamental assumption in the adiabatic picture \cite{Ammosov,Delone}, revealing the non-quasi-static nature of the tunneling process in strong laser fields. Our method is also applicable for molecules, in which the multielectron interaction or the coherent interaction between different orbitals are expected to play important roles \cite{MHe_prl2018,Akagi}. Those interactions during ionization should leave their fingerprint on the momentum as the electron exits the tunneling barrier. Thus, our approach offers the possibility to detect and resolve those multielectron or multi-orbital effects in molecules. 

This work is supported by the National Natural Science Foundation of China (Grant Nos. 11722432,  11674116, 11627809, 11622431, and 61475055) and Program for HUST Academic Frontier Youth Team.

\end{document}